\documentclass{aa}
\usepackage{graphics}
\begin{document}
\thesaurus{11.05.2, 11.09.1, 11.09.1, 11.09.4, 11.19.3, 13.19.1 }%
%
%==================================================================%%
%%                                                                  %%
%%                                                                  %%
%%                                                                  %%
%%                      A S T R O N O M Y                           %%
%%                                                                  %%
%%                           AND                                    %%
%%                                                                  %%
%%                  A S T R O P H Y S I C S                         %%
%%                                                                  %%
%%                                                                  %%
%%        LaTeX Support                             Version 2.10    %%
%%                                                                  %%
%%==================================================================%%
%
%  Abbreviations
%
\def\etal {et al.}
\def\ie {i.\,e.}
\def\etseq {{\em et seq.}}
\def\vs {{it vs.}}
\def\perse {{it per se}}
\def\adhoc {{\em ad hoc}}
\def\eg {e.\,g.}
\def\etc {etc.}
\def\ccpers {\hbox{${\rm cm}^3{\rm s}^{-1}$}}
\def\DEGR {\hbox{$^{\circ }$}}
\def\vlsr {\hbox{${v_{\rm LSR}}$}}
\def\vel {\hbox{${v_{\rm LSR}}$}}
\def\vhel {\hbox{${v_{\rm HEL}}$}}
\def\delv {\hbox{$\Delta v_{1/2}$}}
\def\dvel {\hbox{$\Delta v_{1/2}$}}
\def\TL {$T_{\rm L}$}
\def\TC {$T_{\rm c}$}
\def\TEX {$T_{\rm ex}$}
\def\TMB {$T_{\rm MB}$}
\def\TKIN {$T_{\rm kin}$}
\def\TREC {$T_{\rm rec}$}
\def\TSYS {$T_{\rm sys}$}
\def\TVIB {$T_{\rm vib}$}
\def\TROT {$T_{\rm rot}$}
\def\TDUST {$T_{\rm d}$}
\def\TASTAR {$T_{\rm A}^{*}$}
\def\TVIBST {$T_{\rm vib}^*$} 
\def\TB {$T_{\rm B}$}
\def \la{\mathrel{\mathchoice   {\vcenter{\offinterlineskip\halign{\hfil
$\displaystyle##$\hfil\cr<\cr\sim\cr}}}
{\vcenter{\offinterlineskip\halign{\hfil$\textstyle##$\hfil\cr
<\cr\sim\cr}}}
{\vcenter{\offinterlineskip\halign{\hfil$\scriptstyle##$\hfil\cr
<\cr\sim\cr}}}
{\vcenter{\offinterlineskip\halign{\hfil$\scriptscriptstyle##$\hfil\cr
<\cr\sim\cr}}}}}
\def \ga{\mathrel{\mathchoice   {\vcenter{\offinterlineskip\halign{\hfil
$\displaystyle##$\hfil\cr>\cr\sim\cr}}}
{\vcenter{\offinterlineskip\halign{\hfil$\textstyle##$\hfil\cr
>\cr\sim\cr}}}
{\vcenter{\offinterlineskip\halign{\hfil$\scriptstyle##$\hfil\cr
>\cr\sim\cr}}}
{\vcenter{\offinterlineskip\halign{\hfil$\scriptscriptstyle##$\hfil\cr
>\cr\sim\cr}}}}}
\def\RZWCO {${\cal R}_{2/1}^{\rm C^{18}O}$}
\def\RDRCO {${\cal R}_{3/2}^{\rm C^{18}O}$}
\def\RMSIO {${\cal R}_{5/2}^{\rm ^{28}SiO}$}
\def\RISOSIO {${ r }_{28/29}^{\rm SiO}$}
\def\RISOZSIO {${ r}_{29/30}^{\rm SiO}$}
\def\H0 {$H_{\rm o}$}
\def\mic {$\mu\hbox{m}$}
\def\micro {\mu\hbox{m}}
\def\SDOZ {\hbox{$S_{12\mu \rm m}$}}
\def\STWE {\hbox{$S_{25\mu \rm m}$}}
\def\SSIX {\hbox{$S_{60\mu \rm m}$}}
\def\SHUN {\hbox{$S_{100\mu \rm m}$}}
\def\solmass {\hbox{M$_{\odot}$}}
\def\solum {\hbox{L$_{\odot}$}}
\def\irlum {\hbox{$L_{\rm IR}$}}
\def\ohlum {\hbox{$L_{\rm OH}$}}
\def\blum {\hbox{$L_{\rm B}$}}
\def\numd {\hbox{$n\,({\rm H}_2$)}}                   
\def\rhounit {$\hbox{M}_\odot\,\hbox{pc}^{-3}$}
\def\kms {\hbox{${\rm km\,s}^{-1}$}}
\def\kmsyr {\hbox{${\rm km\,s}^{-1}\,{\rm yr}^{-1}$}}
\def\kmsmpc {\hbox{${\rm km\,s}^{-1}\,{\rm Mpc}^{-1}$}} 
\def\Kkms {\hbox{${\rm K\,km\,s}^{-1}$}}
\def\percc {$\hbox{{\rm cm}}^{-3}$}    %cm-3
\def\cmsq  {$\hbox{{\rm cm}}^{-2}$}    %cm-2
\def\cmsix  {$\hbox{{\rm cm}}^{-6}$}  %cm-6
\def\arcsec {\hbox{$^{\prime\prime}$}}
\def\arcmin {\hbox{$^{\prime}$}}
\def\ffam {\hbox{$\,.\!\!^{\prime}$}}
\def\ffas {\hbox{$\,.\!\!^{\prime\prime}$}}
\def\ffM {\hbox{$\,.\!\!\!^{\rm M}$}}
\def\ffm {\hbox{$\,.\!\!\!^{\rm m}$}}
\def\ffs {\hbox{$\,.\!\!^{\rm s}$}}
\def\ffd {\hbox{$\,.\!\!^{\circ}$}}
\def\HI  {\hbox{HI}}
\def\HII {\hbox{HII}}
%
%   Greek and abbreviations for radio recomb lines etc
%
\def \AL {$\alpha $}    % gr. alpha
\def \BE {$\beta $}     % gr. beta
\def \GA {$\gamma $}    % gr. gamma
\def \DE {$\delta $}    % gr. delta
\def \EP {$\epsilon $}  % gr. epsilon
\def \alde {($\Delta \alpha ,\Delta \delta $)}
\def \MU {$\mu $}       % gr. mue
\def \TAU {$\tau $}     % gr. tau
\def \tapp {$\tau _{\rm app}$}
\def \tuns {$\tau _{\rm uns}$}
\def \RH {\hbox{$R_{\rm H}$}}         % OH main line ratio
\def \RT {\hbox{$R_{\rm \tau}$}}      % OH main tau  ratio
\def \BN  {\hbox{$b_{\rm n}$}}        % bn
\def \BETAN {\hbox{$\beta _n$}}       % beta factor
\def \TE {\hbox{$T_{\rm e}$}}         % Electron Temp.
\def \NE {\hbox{$N_{\rm e}$}}         % Electron Dens.
% molecules
%
\def\MOLH {\hbox{${\rm H}_2$}}                    %H2
\def\HDO {\hbox{${\rm HDO}$}}                     %HDO
\def\AMM {\hbox{${\rm NH}_{3}$}}                  %NH3
\def\NHTWD {\hbox{${\rm NH}_2{\rm D}$}}           %NH2D
\def\CTWH {\hbox{${\rm C_{2}H}$}}                 %C2H
\def\TCO {\hbox{${\rm ^{12}CO}$}}                 %12CO
\def\CEIO {\hbox{${\rm C}^{18}{\rm O}$}}          %C18O
\def\CSEO {\hbox{${\rm C}^{17}{\rm O}$}}          %C17O
\def\CTHFOS {\hbox{${\rm C}^{34}{\rm S}$}}        %C34S
\def\THCO {\hbox{$^{13}{\rm CO}$}}                %13CO
\def\WAT {\hbox{${\rm H}_2{\rm O}$}}              %H2O
\def\WATEI {\hbox{${\rm H}_2^{18}{\rm O}$}}       %H218O
\def\CYAN {\hbox{${\rm HC}_3{\rm N}$}}            %HC3N
\def\CYACFI {\hbox{${\rm HC}_5{\rm N}$}}          %HC5N
\def\CYACSE {\hbox{${\rm HC}_7{\rm N}$}}          %HC7N
\def\CYACNI {\hbox{${\rm HC}_9{\rm N}$}}          %HC9N
\def\METH {\hbox{${\rm CH}_3{\rm OH}$}}           %CH3OH
\def\MECN {\hbox{${\rm CH}_3{\rm CN}$}}           %CH3CN
\def\METAC {\hbox{${\rm CH}_3{\rm C}_2{\rm H}$}}  %CH3C2H
\def\CH3C2H {\hbox{${\rm CH}_3{\rm C}_2{\rm H}$}} %CH3C2H
\def\FORM {\hbox{${\rm H}_2{\rm CO}$}}            %H2CO
\def\MEFORM {\hbox{${\rm HCOOCH}_3$}}             %HCOOCH3
\def\THFO {\hbox{${\rm H}_2{\rm CS}$}}            %H2CS
\def\ETHAL {\hbox{${\rm C}_2{\rm H}_5{\rm OH}$}}  %C2H5OH
\def\CHTHOD {\hbox{${\rm CH}_3{\rm OD}$}}         %CH3OD
\def\CHTDOH {\hbox{${\rm CH}_2{\rm DOH}$}}        %CH2DOH
\def\CYCP {\hbox{${\rm C}_3{\rm H}_2$}}           %C3H2
\def\CTHHD {\hbox{${\rm C}_3{\rm HD}$}}           %C3HD
\def\HTCN {\hbox{${\rm H^{13}CN}$}}               %H13CN
\def\HNTC {\hbox{${\rm HN^{13}C}$}}               %HN13C
\def\HCOP {\hbox{${\rm HCO}^+$}}                  %HCO+
\def\HTCOP {\hbox{${\rm H^{13}CO}^{+}$}}          %H13CO+
\def\NNHP {\hbox{${\rm N}_2{\rm H}^+$}}           %N2H+
\def\CHTHP {\hbox{${\rm CH}_3^+$}}                %CH3+
\def\CHP {\hbox{${\rm CH}^{+}$}}                  %CH+
\def\ETHCN {\hbox{${\rm C}_2{\rm H}_5{\rm CN}$}}  %C2H5CN
\def\DCOP {\hbox{${\rm DCO}^+$}}                  %DCO+
\def\HTHP {\hbox{${\rm H}_{3}^{+}$}}              %H3+ 
\def\HTWDP {\hbox{${\rm H}_{2}{\rm D}^{+}$}}      %H2D+
\def\CHTWDP {\hbox{${\rm CH}_{2}{\rm D}^{+}$}}    %CH2D+
\def\CNCHPL {\hbox{${\rm CNCH}^{+}$}}             %CNCH+
\def\CNCNPL {\hbox{${\rm CNCN}^{+}$}}             %CNCN+
%
% Abbreviations for T. Wiklind article
%
\def\In {\hbox{$I^{n}(x_{\rm k},y_{\rm k},u_{\rm l}$})}
\def\Iobs {\hbox{$I_{\rm obs}(x_{\rm k},y_{\rm k},u_{\rm l})$}}
\def\Ingl {I^{n}(x_{\rm k},y_{\rm k},u_{\rm l})}
\def\Iobsgl {I_{\rm obs}(x_{\rm k},y_{\rm k},u_{\rm l})}
\def\Pbgl {P_{\rm b}(x_{\rm k},y_{\rm k}|\zeta _{\rm i},\eta _{\rm j})}
\def\Pbgm {P(x_{\rm k},y_{\rm k}|r_{\rm i},u_{\rm l})}
\def\Pbgn {P(x,y|r,u)}
\def\Pugm {P_{\rm u}(u_{\rm l}|w_{\rm ij})}
\def\Pdem {P_{\rm b}(x,y|\zeta (r,\theta ),\eta (r,\theta ))} 
\def\Pden {P_{\rm u}(u,w(r,\theta ))}
\def\greekgl {(\zeta _{\rm i},\eta _{\rm j},u_{\rm l})}
\def\greekg1 {(\zeta _{\rm i},\eta _{\rm j})}
\title{The discovery of a gas-rich bar in UGC\,2855:\\
A galaxy in a pre-starburst phase? }
% \subtitle{ }
%
\author{S.~H\"{u}ttemeister\inst{1,2}, S.~Aalto\inst{3,4}, 
W.\ F.\ Wall\inst{5} }
\offprints{S. H{\"u}ttemeister, RAIUB, huette@astro.uni-bonn.de}
\institute{
 Radioastronomisches Institut der Universit\"{a}t Bonn,
 Auf dem H\"{u}gel 71, D - 53121 Bonn, Germany
\and
 Harvard-Smithsonian Center for Astrophysics,
 60 Garden Street, Cambridge, MA 02138, U.S.A.
\and
 Onsala Space Observatory, S - 43992 Onsala, Sweden
\and
 California Institute of Technology, Pasadena, CA 91125, U.S.A.
\and
 INAOE, 72000 Puebla, Mexico }
\date{14 December 1998  / March 9 1999}
\titlerunning{A gas-rich bar in UGC\,2855 }
%\authorrunning{ }
\maketitle
\begin{abstract}
We present the first interferometric CO observations of the barred late-type
galaxy UGC\,2855 and its companion UGC\,2866. UGC\,2855 is shown to belong
to the rare class of galaxies with a long ($\sim 8$\,kpc) continuous 
molecular bar. The velocity field along the bar is dominated by solid-body
rotation and shows few perturbations. This, together with an almost constant
and low $^{12}$CO/$^{13}$CO line intensity ratio along the bar and in the
center as well as only weak H$\alpha$ emission, indicate that the gas in the 
bar is not subjected to strong shocks, but surprisingly quiescent. In the 
central $3''$, a high velocity feature consisting of a number of Giant 
Molecular Associations, is identified. We discuss possible scenarios of
the nature of this structure in connection with the question of the 
presence of an Inner Lindblad Resonance (ILR) in the bar. We suggest that
the bar of UGC\,2855 is a young object, possibly has no ILR, and that 
the mass concentration toward the center of the galaxy is just beginning. 
In contrast, the companion UGC\,2866 experiences a strong starburst. 
We compare the properties of the gas in the bar of UGC\,2855 to those
we find in NGC\,7479. Despite superficial parallels, the two bars are 
very dissimilar objects in terms of $^{12}$CO/$^{13}$CO line ratio, velocity
field and H$\alpha$ activity. NGC\,7479 exhibits starburst characteristics,
while we argue that UGC\,2855 is in a pre-burst stage. 
\end{abstract} 
\keywords{Galaxies: evolution - galaxies: individual: UGC\,2855, UGC\,2866,
 - galaxies: ISM - galaxies: starburst - radio lines: galaxies  }
\section{Introduction}
Barred galaxies may be the rule rather than the exception: It seems 
possible that as many as two thirds of all galaxies have bars (e.g.\
Noguchi 1996). However, most of these stellar bars are almost devoid
of molecular gas. Long, continuous (i.e.\ non-nuclear) gas-rich 
bars that are roughly aligned with the stellar bar seem to be rare
phenomena (e.g.\ Friedli 1998). Most prominent among the few known examples 
are the bars of NGC\,7479 (total length $\sim 10$\,kpc, Sempere et al.\ 1995b, 
Quillen et al.\ 1995, Aalto et al.\ 1998), NGC\,1530 (total length more than 
20\,kpc, Downes et al.\ 1996) and NGC\,2903 (total length $\sim 5.5$\,kpc, 
Jackson et al.\ 1991). Centrally concentrated 
distributions of molecular gas are far more common in barred galaxies; 
a great variety of different morphologies can be found, ranging from 
circumnuclear rings and `twin peak' distributions to central peaks 
(e.g.\ Kenney et al.\ 1992, Kenney 1996).

\begin{table*}
\caption{\label{prop} Properties of UGC\,2855 and UGC\,2866}
\begin{tabular}{lrrrrrrr}
Name & \multicolumn{1}{c}{R.A. (1950.0)$^{\rm a)}$} 
& \multicolumn{1}{c}{Dec (1950.0)$^{\rm a)}$} 
& Type & $m_b$ & Size (blue) & Inclination$^{\rm b)}$ 
& $L_{\rm FIR}$ \\
\hline \\
UGC\,2855 & $03^{\rm h}\ 43^{\rm m}\ 13.7^{\rm s}$ & $69^{\circ}\ 58'\ 44''$ &
SBc$^{\rm c)}$ & 13\ffm 5$^{\rm c)}$  
& 4\ffam 6 $\times$ 2\ffam 2$^{\rm d)}$ & $\sim 60^{\circ}$ 
& $4.6 \cdot 10^{10}$\solum$^{\rm e)}$ \\
UGC\,2866 & $03^{\rm h}\ 45^{\rm m}\ 07.2^{\rm s}$ & $69^{\circ}\ 56'\ 34''$ &
Early S?/H\,{\sc II}$^{\rm f)}$ & 15\ffm 5$^{\rm d)}$  
& 1\ffam 1 $\times$ 0\ffam 9$^{\rm d)}$ & $30^{\circ} - 35^{\circ}$ 
& $4.9 \cdot 10^{10}$\solum$^{\rm g)}$ \\
\hline 
\end{tabular} \\
a): Radio continuum peak position (Condon et al.\ (1996) \\
b): calculated from the axes ratio \\
c): from the RC3 catalogue (de Vaucouleurs et al.\ 1991) \\
d): from the UGC catalogue (Nilson 1973) \\
e): from Condon et al.\ (1996) \\
f): from Cohen \& Volk 1989 \\
g): from Zenner \& Lenzen 1993 
\end{table*}

While the necessity of bars in general in triggering nuclear activity or
enhanced star formation, up to starbursts, is somewhat controversial 
(Hawarden et al.\  1996, Ho et al.\ 1997), they definitely provide a 
mechanism of funnelling gas from the outer parts of a galaxy to its 
central region (Athanassoula 1992). As a result of tidal strain, shocks 
and cloud-cloud collisions that should occur in the non-axisymmetric 
potential of a bar, gaseous bars also are a laboratory to study diffuse 
and possibly unbound gas (e.g.\ Das \& Jog 1995). This is a type of 
interstellar medium (ISM) likely to be important in the bulge and central 
regions of many galaxies including the Milky Way (Dahmen et al.\ 1998), 
the properties of which are largely unexplored. 

Since the triggering of bars is thought to be due to galaxy interactions 
in many, though not all, cases, good candidates for gaseous, `active' bars 
are galaxies in the early stages of interaction, when the gas flow along 
the bar has just started. We therefore launched a project in which
we search distant pairs of galaxies known to contain molecular gas and a
stellar bar for gas along the bar. 

In this paper, we report the discovery of one of the longest continuous 
molecular bars known, extending over $\sim 8$\,kpc, in the galaxy
UGC\,2855. UGC\,2855 and its companion UGC\,2866, at a projected
distance of 10$'$, have never been studied individually but only as parts 
of surveys (in molecular gas by Sofue et al.\ 1997, Elfhag et al.\ 1996, 
Aalto et al.\ 1995, Young et al.\ 1995). Both of them 
are, however, FIR-bright IRAS galaxies containing a large amount of 
molecular gas (Aalto et al.\ 1995). We sum up their few known properties 
in Table\,\ref{prop}. Following Condon et al.\ (1996), a distance of 20\,Mpc 
to both galaxies is adopted throughout this paper. Thus, an angular size of 
$5''$ corresponds to 480\,pc on a linear scale, and UGC\,2855 and UGC\,2866
have a (projected) distance to each other of $\sim 60$\,kpc. Sofue et al.\ 
(1997), who included UGC\,2855 in their sample of galaxies for which they
obtained rotation curves from CO single dish observations, derive a distance 
of 28.77\,Mpc from the Tully-Fisher relation. If this distance is correct, 
the length of the bar in UGC\,2855 rises to $\sim$\,10\,kpc.

\section{Observations}

\subsection{Interferometric $^{12}$CO observations}

Interferometric $^{12}$CO $1\to 0$ observations were obtained at the Owens 
Valley Radio Observatory (OVRO) millimeter array. The six 10.4\,m telescopes
are equipped with SIS receivers providing system temperatures of typically
$\sim 500$\,K (single sideband). Details of the observations are given in
Table \ref{log}. With natural weighting, we obtained a roughly circular 
beam of size $5\ffas 4 \times 5\ffas 2$ (PA 46$^{\circ}$ and sensitivity
to extended structure up to 17$''$. The quasar 2345--167 was 
observed as a phase and amplitude calibrator every $\sim 15$\,minutes. 
Passband calibration was done using stronger sources, usually 0528+134 
or 3C\,454.4. These sources, calibrated against the planets Neptune and 
Uranus, were also taken for flux calibration. The tracks were calibrated 
using software developed for the OVRO array. After calibration, the data 
were inspected for quality, and low-coherence data points were edited 
out on a baseline-to-baseline basis. 

\begin{table*}
\caption{\label{log} Summary of the interferometric observations}
\begin{tabular}{rrlllll}
\multicolumn{1}{c}{Source and Line} 
& \multicolumn{1}{c}{Date} 
& Field Center & $uv$-coverage & Number of Fields & Primary Beam  
& Config.$^{a)}$ \\
\hline \\
UGC\,2855; $^{12}$CO($1\to 0)$ & 5/96, 9/96 & 
$\alpha = 03^{\rm h}\ 43^{\rm m}\ 12.4^{\rm s}$ &
6k$\lambda$ -- 42\,k$\lambda$ & 3; spaced by 
& 65$''$ & 3: C,L,H \\
 & 5/97 & $\delta = 69^{\circ}\ 58'\ 37''$ & (15m -- 115m) & 45$''$ along bar &
 & \\
 $^{13}$CO$(1\to 0)$ & 5/98  & 
$\alpha = 03^{\rm h}\ 43^{\rm m}\ 12.4^{\rm s}$ &
5.7k$\lambda$ -- 82\,k$\lambda$ & 3; spaced by 
& 65$''$ & 1: L \\
 &  & $\delta = 69^{\circ}\ 58'\ 37''$ & (15m -- 230m) & 45$''$ along bar &
 & \\
\hline
UGC\,2866; $^{12}$CO$(1\to 0)$ & 5/96 & 
$\alpha = 03^{\rm h}\ 45^{\rm m}\ 7.9^{\rm s}$ &
6k$\lambda$ -- 44\,k$\lambda$ & 1
& 65$''$ & 1: L \\
 & 5/97 & $\delta = 69^{\circ}\ 56'\ 30''$ & (15m -- 230m) &  &
 & \\
\hline 
\end{tabular} \\
$a)$: The array configurations at OVRO are labelled C (compact), L (low
resolution) and H (high resolution) 
\end{table*}

For a beam size of $5\ffas 3$, a brightness temperature (\TB ) of 1\,K  
corresponds to 0.304\,Jy beam$^{-1}$ at a wavelength of 2.6\,mm. The 
spectral resolution of our data is 4\,MHz or 10.04\,\kms, and the total 
velocity range covered by the autocorrelator is 1120\,\kms , centered 
on \vlsr = 1200\,\kms . 

Imaging was done using the NRAO AIPS package. To cover the long bar of 
UGC\,2855 fully, we obtained a 3-field mosaic with a spacing of $45''$ 
between the pointing centers. The fields were CLEANed individually and 
combined linearly, correcting for the attenuation of the primary beam,
using the AIPS task LTESS.  

\begin{figure*}
\resizebox{!}{21cm}{\includegraphics{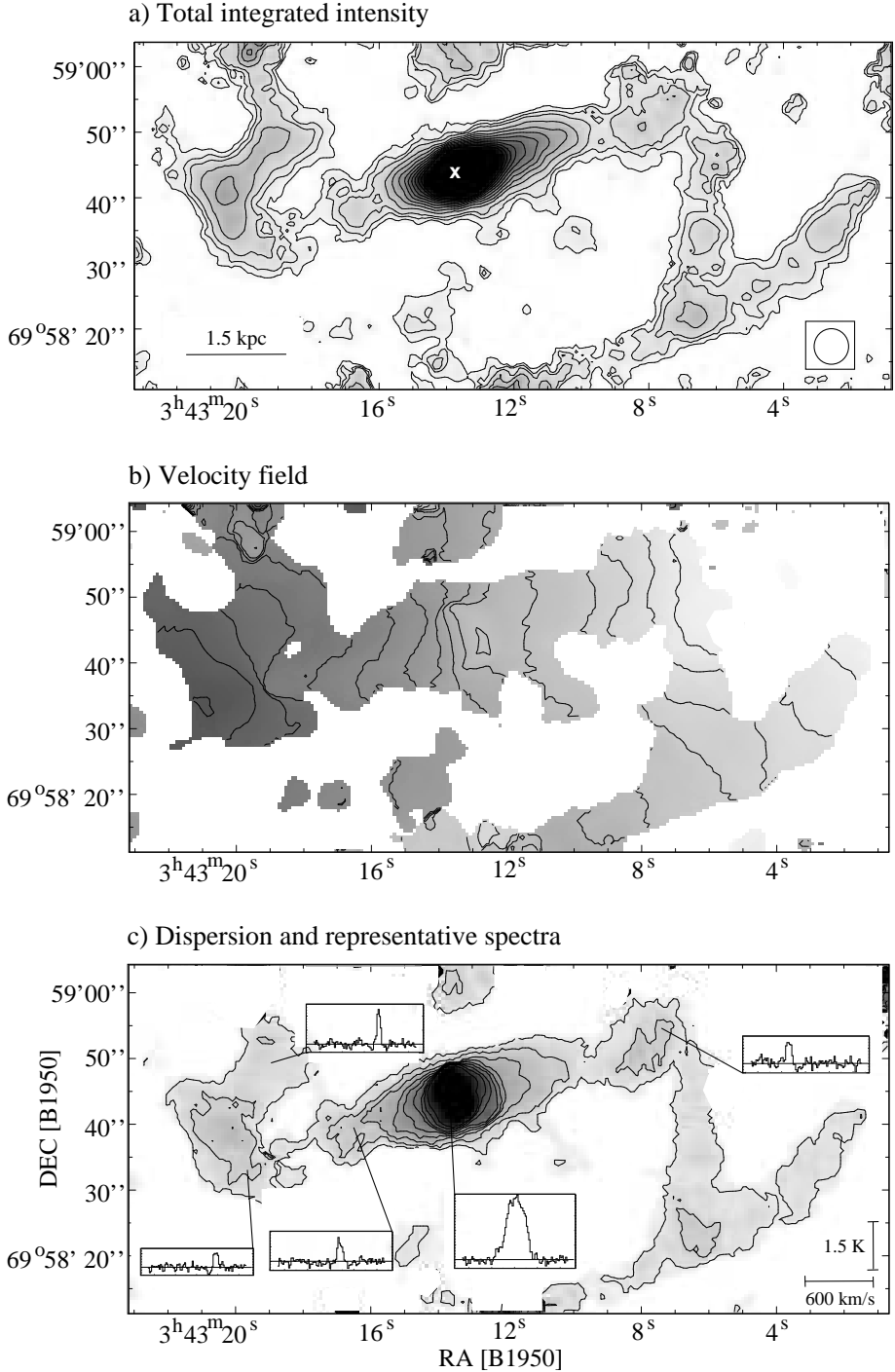}}
\caption{\label{mom} The bar of UGC\,2855: a) Total integrated
intensity. The contours start at 13.8 K\,\kms\ (the $2\sigma$ 
level, corresponding to 4.2\,Jy\,beam$^{-1}$\,\kms ), and increase in 
steps of $2\sigma$. The white cross marks the position of the 
radio continuum peak.
b) Velocity field. The contours range from 1050\,\kms\ to 1400\,\kms\ in steps
of 25\,\kms , with darker color in the underlying grey scale indicating 
redshifted emission.
c) Velocity dispersion. The contours range from 10\,\kms\ to 50\,\kms\ 
in steps of 5\,\kms . The highest value (black) is 73\,\kms . The changes 
in line shape are illustrated by spectra taken at the center and 
different offsets along the bar. These are centered on 1200\,\kms ; their 
scale is given in the lower right corner of the figure.}
\end{figure*}

To analyse the innermost structure (see Sect.\
4.1), the beam size could be reduced to $3\ffas 4 \times 3\ffas 3$ 
(PA 78$^{\circ}$) applying robust weighting (Briggs 1995), while a beam size 
of $2\ffas 1 \times 2\ffas 0$ (PA --19$^{\circ}$) was reached when only the 
longer baseline data from the high resolution track were used. This
(naturally weighted, since the signal-to-noise in the robustly weighted maps 
is too low) map includes baselines from 230\,m to 35\,m length, which means 
that structures larger than $\sim 7''$ cannot be imaged. The increased 
resolution also goes along with a loss in sensitivity: The rms noise per
channel increases from 0.11\,K \TB\ (34\,mJy\,beam$^{-1}$)  in the naturally
weighted map to  0.43\,K (52\,mJy\,beam$^{-1}$) in the robustly weighted and 
1.34\,K (62\,mJy\,beam$^{-1}$) in the high-resolution-only maps. 

The interferometric $^{12}$CO observations of UGC\,2866 are summed up in
Table \ref{log}. The naturally-weighted map of UGC\,2866 has a beam size 
of $4\ffas 2 \times 4\ffas 1$ (PA --68$^{\circ}$) and an rms noise of 
0.20\,K (38\,mJy\,beam$^{-1}$). The velocity resolution, velocity range and 
central velocity of these observations were the same as those of UGC\,2855. 

\subsection{Interferometric $^{13}$CO observations}

Data in the $^{13}$CO $1\to 0$ transition of the same
three fields in UGC\,2855 (see Table \ref{log}) that were mapped in 
$^{12}$CO were taken at 
OVRO in June 1998 under excellent weather conditions, using the same
spectrometer configuration as for $^{12}$CO. The calibration and mapping 
procedure was identical to the $^{12}$CO measurements. The (naturally weighted) 
synthesised beam has a size of $5\ffas 7 \times 4\ffas 5$ (PA $-16^{\circ}$), 
and the sensitivity reached is 0.09\,K \TB\  (23\,mJy\,beam$^{-1}$).

\subsection{Single-dish observations of UGC\,2855} 

Single dish $^{12}$CO $1\to 0$ and $^{13}$CO $1\to 0$ data of UGC\,2855 were
obtained with the 20\,m telescope at Onsala Space Observatory in December 
1997 and January 1998. We observed 7 spectra along the bar in $^{12}$CO, 
spaced by 15$''$ along the bar, and 5 positions in $^{13}$CO 
(see Fig.\,\ref{oso}). The central 
position ($\alpha = 03^{\rm h}\ 43^{\rm m}\ 13.6^{\rm s}; \delta = 
69^{\circ}\ 58'\ 44'' (1950.0) $ almost exactly coincided with 
the interferometric CO emission peak. The typical system 
temperature was 500\,K -- 600\,K for $^{12}$CO and 200\,K -- 300\,K 
for $^{13}$CO. The beam size is $33''$, and the spectra have been smoothed 
to a velocity resolution of $\sim 10$\,\kms . The Onsala spectra do not
fully sample the entire area of the OVRO interferometer map and have
pointing uncertainties of $\sim 5''$, preventing us from combining the
single-dish data with the interferometry data into a single map. 

\subsection{H$\alpha$}  

An H$\alpha$ image of UGC\,2855 and UGC\,2866 was taken using HoLiCam (Reif
et al.\ 1995) at the 1.06\,m telescope at Hoher List Observatory in 
October 1997. The conditions were not photometric, and the image 
provides relative information on fluxes only. However, no other 
deep H$\alpha$ imaging of these galaxies is available at this time. The 
image was processed using the IRAF package.

\section{Results}

\subsection{UGC\,2855}

UGC\,2855 is a large spiral galaxy, almost the same size as the Milky Way,
classified as SBc (RC3, de Vaucouleurs et al.\ 1991). At 20\,Mpc distance, 
its linear diameter to the $D_{25}$ radius is 25\,kpc. Assuming a Galactic 
extinction in the $B$ band of $1\ffm 69$ (Burstein \& Heiles 1984), the 
absolute magnitude of UGC\,2855 is ${\rm M}_B = -19\ffm 7$, also comparable
to the Milky Way.  As it is typical for barred spirals, its main spiral 
arms start at the ends of the bar, and strong molecular emission can be 
traced in both the bar and the arms.

\begin{figure*}
\resizebox{\hsize}{!}{\includegraphics{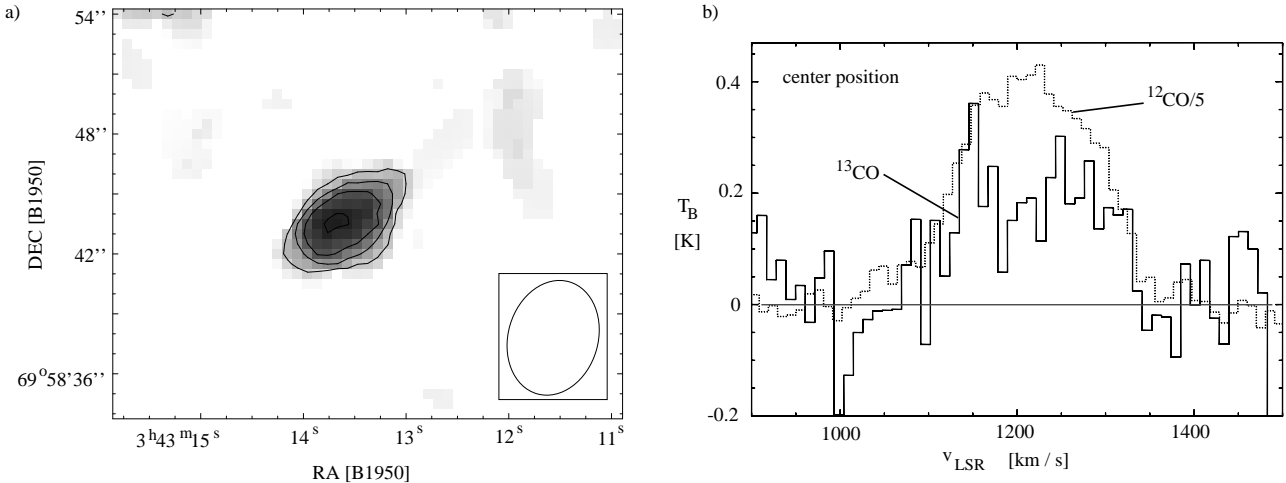}}
\caption{ \label{13co} OVRO observations of $^{13}$CO in UGC2855. a) Total
integrated intensity. The contours start at the 2$\sigma$ level, 
10\,K\,\kms\ (2.6\,Jy\,beam$^{-1}$\,\kms ), and are spaced by 2$\sigma$. b) 
Comparison between the central $^{13}$CO $1\to 0$ and $^{12}$CO $1\to 0$
(divided by 5 for better comparability) spectra. The width and, within the
noise, shape of the two spectra are identical. Due to the low signal-to-noise
in the $^{13}$CO spectrum, the suggested rise in $^{12}$CO/$^{13}$CO ratio 
at the center of the line is not statistically significant beyond the 
1.5$\sigma$ level.} 
\end{figure*} 

\subsubsection{The distribution of $^{12}$CO}

We present the total integrated intensity map, the velocity field and the
map of the velocity dispersion from our interferometric $^{12}$CO 
observations in Fig.\,1. From Fig.\,1\,a, it is clear
that the bar in UGC\,2855 is filled with molecular gas over its entire length.
The position of the peak total intergrated intensity at $\alpha = 
03^{\rm h}\ 43^{\rm m}\ 13\ffs 6$ and $\delta = 69^{\circ}\,58' 44''$
agrees almost exactly with the radio continuum peak ( $\alpha = 
03^{\rm h}\ 43^{\rm m}\ 13\ffs 7$ and $\delta = 69^{\circ}\,58' 44''$,
Condon et al.\ 1996). The overall distribution of the molecular gas in the interferometer map is somewhat asymmetric: The northwestern part of the 
bar outside the center is brighter in CO by about a factor of 1.5 than 
its southeastern counterpart. This effect can be explained, at least partially,
by the fact that the interferometer misses more flux in the southeast
of the bar than in the northwest (see Sect. 3.1.2). 

\begin{figure}
\resizebox{\hsize}{!}{\includegraphics{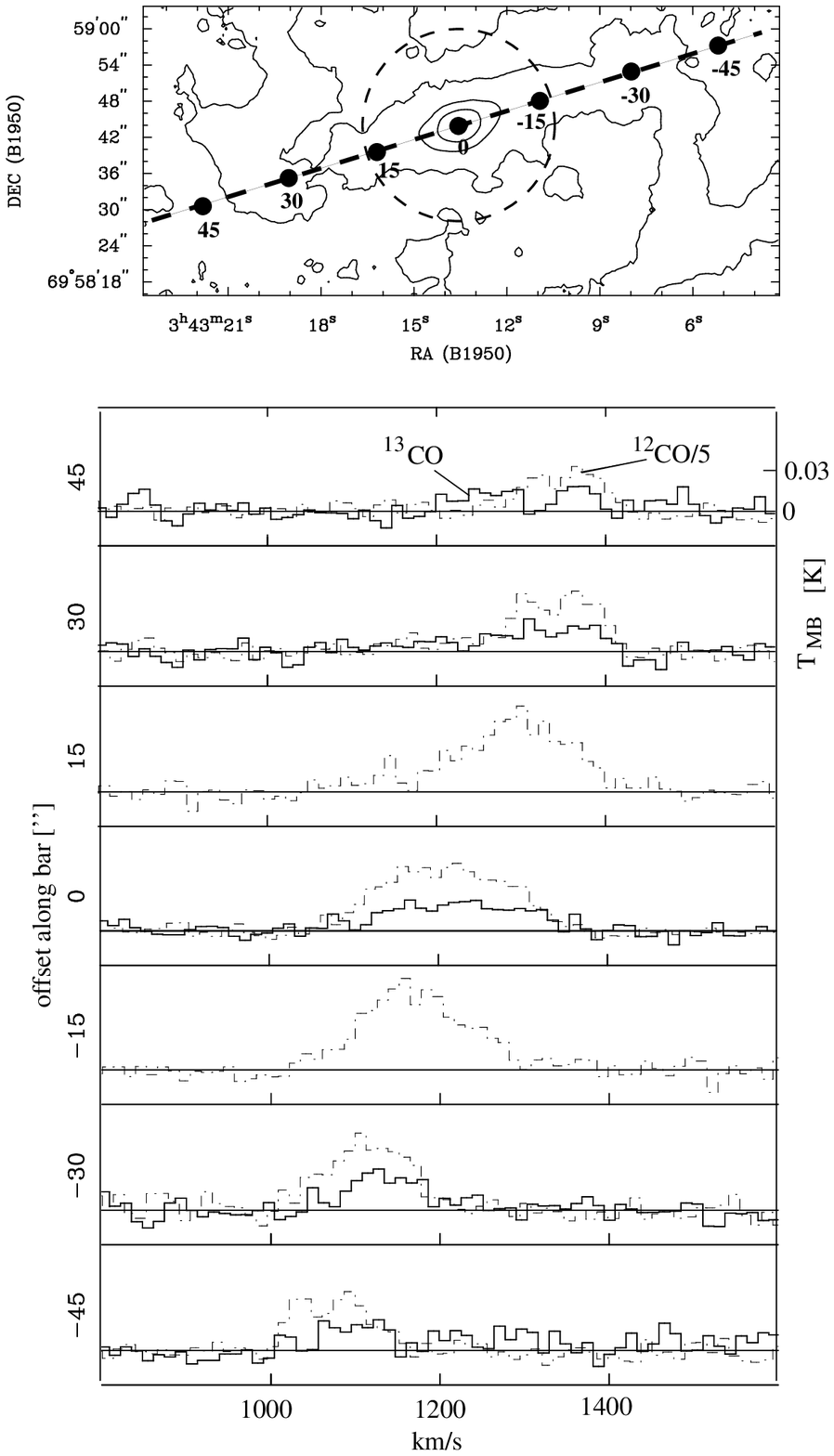}}
\caption{\label{oso} OSO single dish observations of the bar of UGC\,2855.
The finder chart (top panel) identifies the positions of the spectra
with respect to the interferometry map. The OSO FWHM beam is indicated
for the central position. 
$^{13}$CO spectra are shown as a solid line, $^{12}$CO spectra (divided by
5 for better comparability) are shown as dashed lines. The spectra are on a
\TMB\ scale, based on $\eta_{\rm MB} = 0.5$. Note the good agreement
in the line shapes. }
\end{figure}

The bar major axis has a position angle of 110$^{\circ}$ and is thus 
aligned very closely with the line of nodes (the major axis) of the galaxy,
which is at PA $=109^{\circ}$ according to the I-band surface photometry 
of H\'eraudeau \& Simien (1996). This implies that the bar is seen side-on, 
at its maximum length. Assuming this viewing geometry and an inclination 
of $60^{\circ}$, the CO bar can be deprojected, following the formalism 
given by Martin (1995).
The deprojected ratio of bar length to $D_{25}$ diameter is 0.3, while 
we determine the axis ratio of the minor and major bar axis to be 0.37. 
According to the classification of Martinet \& Friedli (1997),
the bar of UGC\,2855 thus qualifies as both `long' and `strong'. 

The velocity field along the bar (panel b in Fig.\,1) appears
regular and suggests a solid body rotation. The isovelocity contours are 
roughly perpendicular to the bar major axis. The meaning of slight 
deviations from this behaviour will be explored further in Sect.\ 4.1.2. 

The peak of the velocity dispersion $\sigma_{\rm v}$ 
(displayed in Fig.\,1, panel c) coincides with the intensity peak. 
Along the bar, the dispersion drops by a factor of about 5 from 
$\sim 50$\,\kms\  to $\sim 10$\,\kms . At the bar ends, $\sigma_{\rm v}$ 
rises slightly to $\sim 15$\,\kms . The narrowness of the lines along the bar 
is also demonstrated by sample spectra in Fig.\,1\,c. 

\subsubsection{Line fluxes and the molecular mass} 

The total flux in the map is $\sim 1200$\,Jy\,\kms . It is evident from 
optical images that the spiral arms extend far beyond our map. Thus, this 
CO flux is a lower limit only and demonstrates that UGC\,2855 
is a gas-rich galaxy. Assuming a galactic `standard' $I$(CO)--$N$(H$_2$)
conversion factor (SCF, $2.3 \cdot 10^{20}$\,cm$^{-2}$ (K\ \kms )$^{-1}$, e.g.\ 
Strong et al.\ 1988), this corresponds to a total H$_2$ mass of 
$\sim 4.5 \cdot 10^9$\,M$_{\odot}$.  
This conversion factor is suspect, especially in the centers of galaxies 
(Dahmen et al.\ 1998). It may vary greatly between 
galaxies and also within a galaxy, depending on the properties of the 
molecular gas, e.g.\ the amount of diffuse gas, and on the metallicity 
(Arimoto et al.\ 1996). We give a rough estimate of how far the SCF might
overestimate the molecular mass by comparing it to the result we derive from
the $^{13}$CO line intensity. We assume LTE, optically thin $^{13}$CO 
emission, a kinetic temperature of 20\,K, a $^{12}$CO/$^{13}$CO ratio
of 30 (thought to be typical for centers of galaxies) and a $^{12}$CO/H$_2$
ratio of $10^{-4}$. Then, the H$_2$ column density derived from $^{13}$CO
for the center of UGC\,2855 is lower by a factor of 8 than the column 
density calculated using the SCF. Interestingly, this value is the same 
for the interferometric and the single-dish observations, even though the
absolute values for the column densities are higher for the far smaller
interferometer beam by a factor of more than 10. To force agreement between 
the column density derived from $^{13}$CO and the SCF, we would have to
raise \TKIN\ to 190\,K. For positions $30''$ off the center,
the correction implied by $^{13}$CO is smaller by as much as a factor of two. 
Corroborating evidence from our galaxy as well as external galaxies (e.g.\
Wall et al.\ 1993) confirms that gas column densities determined from 
the SCF and $^{13}$CO generally agree better away from the nuclear 
regions. Of course, the estimates based on $^{13}$CO depends on the 
assumed value of \TKIN : If there is a gradient in \TKIN\ with galactocentric
radius (with hotter gas closer to the nucleus), this effect may be cancelled.
Thus, the assumptions made for the $^{13}$CO emission are vastly 
oversimplified, but the estimate demonstrates the range of uncertainty of 
the molecular masses.

Throughout this paper, we will
give masses based on the SCF to facilitate comparisons with other work and 
also because a determination of  a `correct' conversion factor that goes beyond 
the estimate given above would require detailed information on the 
gas properties not yet available. {\em It should be kept in mind, though, that 
the masses may be lower by almost an order of magnitude.\/} 

As a further complication, the interferometer is likely to filter out flux 
from an extended gas component due to missing zero spacing coverage. We 
find a shallow negative trough surrounding the strongest features in our 
map, indicating that the array may indeed miss some extended emission. 
To get a rough estimate of the amount of missing flux, we have compared the
flux contained in single Onsala beams (see Fig.\,\ref{oso}) to the flux 
in equivalent regions of the interferometer beam, weighted by the shape 
and size of the single
dish beam (assumed to be gaussian). From this, toward the center and the
northwestern part of the bar, we see at least 80\% of the single dish 
dish flux in the interferometer map. For the southeastern part of the
bar, there are indications that a somewhat lower amount of flux is picked
up by the interferometer, but still at least 65\% (and possibly as much as 
80\%) of the single dish flux is seen.  These numbers imply that not much
gas is distributed in a very diffuse component that has little structure 
on scales exceeding $\sim 1.5$\,kpc. However, all
pointing and calibration errors enter in this estimate, which therefore 
has to be considered to be very tentative. Clearly, in the mass estimates,
the error introduced by the missing flux is dominated by the uncertainty
of the conversion factor. 

The total flux the interferometer detects along the bar, including the bar
end regions, but excluding the beginning of the spiral arms, 
is 895\,Jy\,\kms . This corresponds to a molecular mass of $\sim 3 \cdot 
10^9$\,M$_{\odot}$, under the assumptions described above. 
The central beam contains a flux of 137\,Jy\,\kms\ ($\sim 5 \cdot
10^8$\,M$_{\odot}$).

\subsubsection{$^{13}$CO}

The results of the OVRO $^{13}$CO measurements are presented in Fig.\,\ref{13co}.
Panel a shows the total integrated intensity distribution. $^{13}$CO is
only detected toward the center of UGC\,2855. Assuming a constant 
$^{12}$CO/$^{13}$CO line intensity ratio (see below), this is, however,
expected, since the sensitivity in the $^{13}$CO map is not sufficient to 
detect emission away from the intensity peak. The positions of the $^{13}$CO 
peak and the $^{12}$CO peak are identical. To within the noise, the linewidth 
and the line shape of the two isotopomers also agree (Fig.\,\ref{13co}, 
panel b). The observations taken at the OSO 20\,m telescope (Fig.\,\ref{oso}) 
confirm the good agreement between the line shapes of $^{12}$CO and $^{13}$CO.  

\subsubsection{H$\alpha$}

\begin{figure*}
\resizebox{\hsize}{!}{\includegraphics{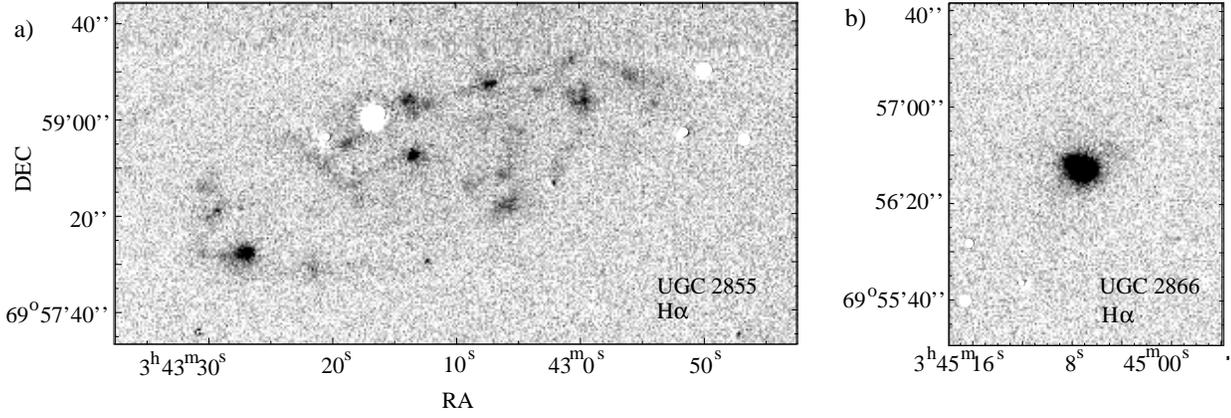}}
\caption{\label{halpha} H\,$\alpha$-images of UGC\,2855 (a) and UGC\,2866
(b), taken with HoLiCam at the 1.0\,m telescope at Observatory Hoher List 
(Bonn University). Continuum emission has been subtracted using a red image.
Thus, bright foreground stars appear as white dots in the images. Both 
images are close-ups from the same CCD frame. The positions of $\sim 30$ 
stars identified in an DSS image have been used to determine the coordinate 
system of the CCD frame. }
\end{figure*}

Figure \ref{halpha} (panel a) shows our H$\alpha$ image of UGC\,2855. Since
the image lacks an absolute scale, absolute H$\alpha$ fluxes or star formation
rates cannot be derived. However, it gives information on the relative 
strength of H$\alpha$ emission and thus star formation activity, in the
center, along the bar and in the spiral arms. H$\alpha$ emission is clearly
detected in the center, but the central peak is not the strongest one in
the galaxy. The peak in the southeastern spiral arm, outside the limits of
our CO map, is almost twice as strong as the central emission, and the 
emission from three more peaks, all at the very edge or outside our CO
mosaic, has roughly the same strength as the central emission. This
demonstrates that, while there is some central activity, the nucleus of 
UGC\,2855 is not (yet?) in a starburst phase. Of course, the nuclear H$\alpha$ 
emission may be subject to more extinction than the emission from the
spiral arms (Phillips 1996). However, starburst galaxies show nuclear 
H$\alpha$ emission that is clearly enhanced over that seen in their disks,
despite having stronger extinction effects in their nuclei than in their
disks (e.g.\ Contini et al.\ 1998, who find that the H$\alpha$ luminosity
in the starburst nuclei of their sample is higher by a factor of $\sim 10$
than in extranuclear H\,II regions; another example is UGC\,2866, see 
Sect.\ 3.2).  

Along the bar, in particular
the southeastern part, faint H$\alpha$ emission is detected. Even though
the subtraction of the background is problematic, we estimate that the
total emission (not corrected for extinction) from the bar and the emission 
from the central region are about equal. 

\subsection{The companion: UGC\,2866}

\begin{figure}
\resizebox{\hsize}{!}{\includegraphics{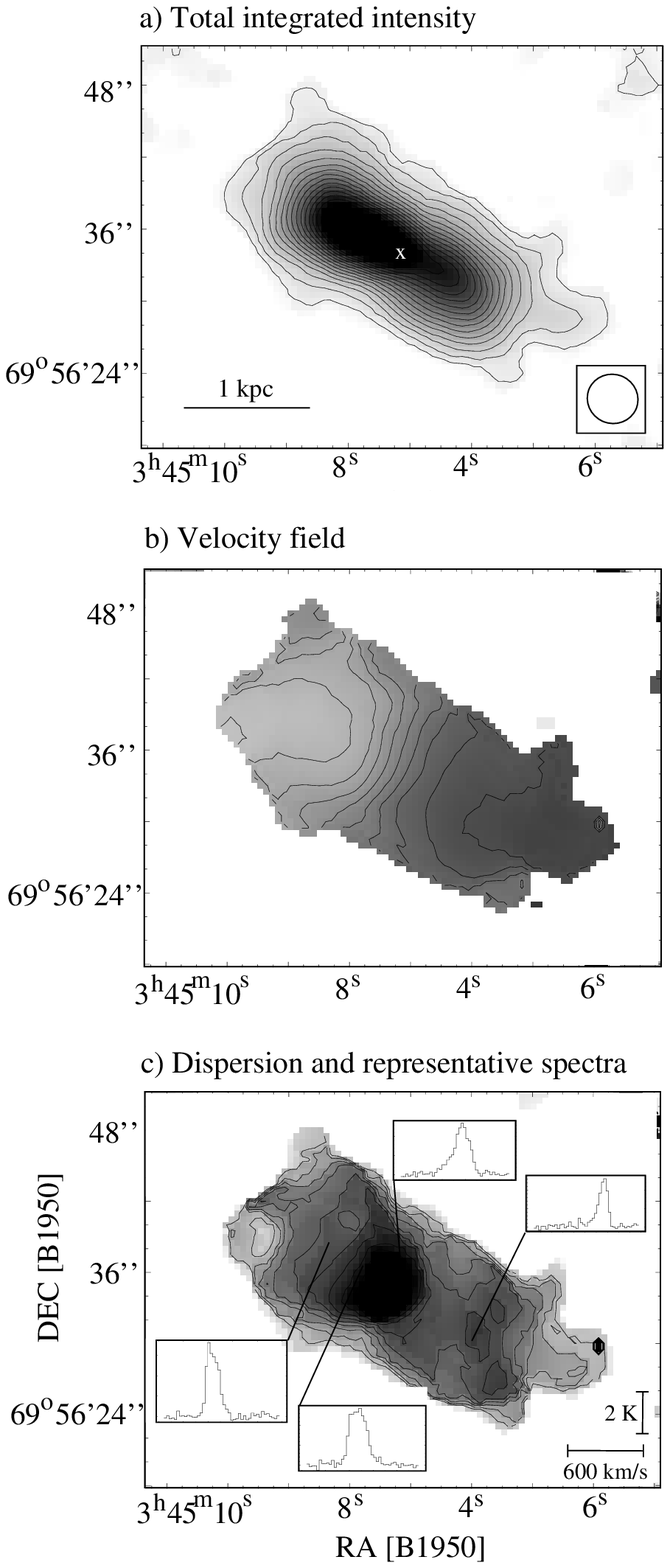}}
\caption{\label{2866mom} UGC\,2866: a) Total integrated
intensity. The contours start at 4.6\,Jy (24.6\,K\,\kms , $2\sigma)$ and 
increase in $2\sigma$ steps. As in Fig.\,\label{mom}, the radiocontinuum
is indicated by a cross. 
b) Velocity field. The contours range from 1125\,\kms to 1450\,kms , in steps
of 25\,\kms, darker color in the greyscale image indicates redshifted emission.
c) Velocity dispersion. The contours range from 20\,\kms to 55\,\kms, 
the dispersion peak (black) is at 75\,\kms . The spectra shown are centered
on 1250\,\kms. Their scale is given in the lower right corner of the panel.}
\end{figure}

The picture presented by UGC\,2866, the smaller, distant companion of 
UGC\,2855, is very different. The classification of this galaxy is not
clear: The UGC catalogue (Nilson 1973) tentatively calls it an early spiral,
while it is (mis)classified as an elliptical by Hau et al.\ 1995. It is one
of only 21 galaxies from the IRAS PSC that are stronger than 1.5\,Jy at both
12$\mu$m and 25$\mu$m with IRAS LRS spectra allowing its identification as 
an H\,{\sc II} galaxy in the sample of Cohen \& Volk (1989). All the 
molecular gas of UGC\,2866 seems to be concentrated in one feature of 
slightly more than 2\,kpc length that has an oval, possibly bar-like 
appearance (Fig.\,\ref{2866mom} a). The emission peaks at $\alpha = 
03^{\rm h}\ 45^{\rm m}\ 7\ffs 7$ and $\delta = 69^{\circ}\,56' 36''$,
unusually far ($3''$) offset from the radio continuum peak at $\alpha = 
03^{\rm h}\ 45^{\rm m}\ 7\ffs 2$ and $\delta = 69^{\circ}\,56' 34''$
(Condon 1996). 

The generally smooth, almost classical `spider'
shape of the velocity field (Fig.\,\ref{2866mom} b) indicates gas that
is predominantly in circular motion, though some deviations, especially 
close to the edges of the structure, are present. The velocity
dispersion is, once more, highest in the center, but the drop along the 
major axis is less extreme than in UGC\,2855. For an offset along the
major axis of UGC\,2855 of $+10''$ and $-10''$, the dispersion drops from the
central peak of 73\,\kms\ to 23\,\kms\ and 19\,\kms\, respectively, while
the drop for corresponding offsets in UGC\,2866 (central peak at 75\,\kms )
is only to 36\,\kms\ and 44\,\kms . 

We detect a total flux of 543 Jy\,\kms\ in our map of UGC\,2866, which,
again assuming a `standard' conversion factor, translates to an H$_2$-mass
of $\sim 2.0 \cdot 10^9$\,M$_{\odot}$, not much smaller than what is found in 
the bar of UGC\,2855. The central beam contains $\sim 3.5 \cdot
10^8$\,M$_{\odot}$ of H$_2$. If the radiocontinuum peak is identified 
with the dynamical center of the galaxy, about 55\% of the CO emission 
arises in the northeastern half, which also contains the intensity maximum.  

The H$\alpha$ emssion from UGC\,2866 (Fig\,\ref{halpha}, panel b, a part
of the same CCD frame as the image of UGC\,2866) is very bright and very 
compact: it extends over only $\sim 8''$ or $\sim 700$\,pc. The intensity of 
the emission is more than 6 times higher than what is found toward the 
center of UGC\,2855. The center position of H$\alpha$ coincides with the
CO peak in UGC\,2866 to better than 1$''$, i.e.\ to within the measuring
uncertainty. The strength and compactness of H$\alpha$ in UGC\,2866 and
the FIR properties (see below) indicate that this galaxy is experiencing 
a nuclear starburst. 

\section{Discussion}

\subsection{Gas surface densities}

In the central beam, the surface density of UGC\,2855 (corrected for an 
inclination of $60^{\circ}$) is $\Sigma_{\rm cen} = 
1200$\,M$_{\odot}$\,pc$^{-2}$. In the inner part of the bar, this value 
drops to $\sim 550$\,M$_{\odot}$\,pc$^{-2}$. UGC\,2866 has a central mass 
surface density of $\Sigma_{\rm cen} = 2400$\,M$_{\odot}$\,pc$^{-2}$ (for
$i = 30^{\circ}$). These numbers (remember, however,
that they are based on the SCF) are high enough to place UGC\,2855 and
especially UGC\,2866 in the range of IR-bright starburst galaxies (e.g.\ 
Scoville 1991), but only UGC\,2866 clearly is a 
starburst. A significantly lower gas mass (as discussed in Sect.\ 3.1.2)
would reduce $\Sigma $; however, a similar correction is likely to be
also necessary for the sample of IR bright starbursts, so that UGC\,2855
and UGC\,2866 would remain in this group.
Starburst properties depend on morphology and dynamics (e.g.\ 
Kenney 1997), but the central concentration found in UGC\,2855 should be a 
fairly favorable environment. We will, however, speculate below that 
UGC\,2855 may be in a preburst phase. 

\subsection{Kinematics of the gas: UGC\,2855}

\begin{figure}
\resizebox{\hsize}{!}{\includegraphics{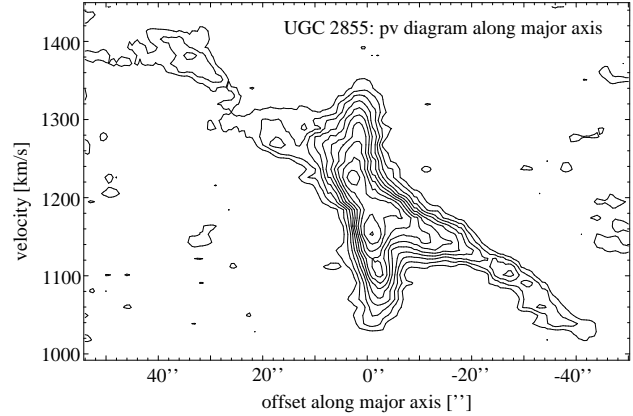}}
\caption{\label{2855pv} Position-velocity diagram of UGC\,2855 along the
major axis, from the naturally weighted data set. The cut is centered on the 
CO integrated emission peak at $\alpha = 03^{\rm h} 43^{\rm m} 13.6^{\rm s};\
\delta = 69^{\circ} 58' 44''\ (1950.0)$. Contours start at the $2\sigma$ level
(0.22\,K\,\TB  or 68\,mJy\,beam$^{-1}$) and range to 2.33\,K in
steps of $2\sigma$.}
\end{figure}

\subsubsection{Rigid rotation and a high velocity feature}

\begin{figure*}
\resizebox{\hsize}{!}{\includegraphics{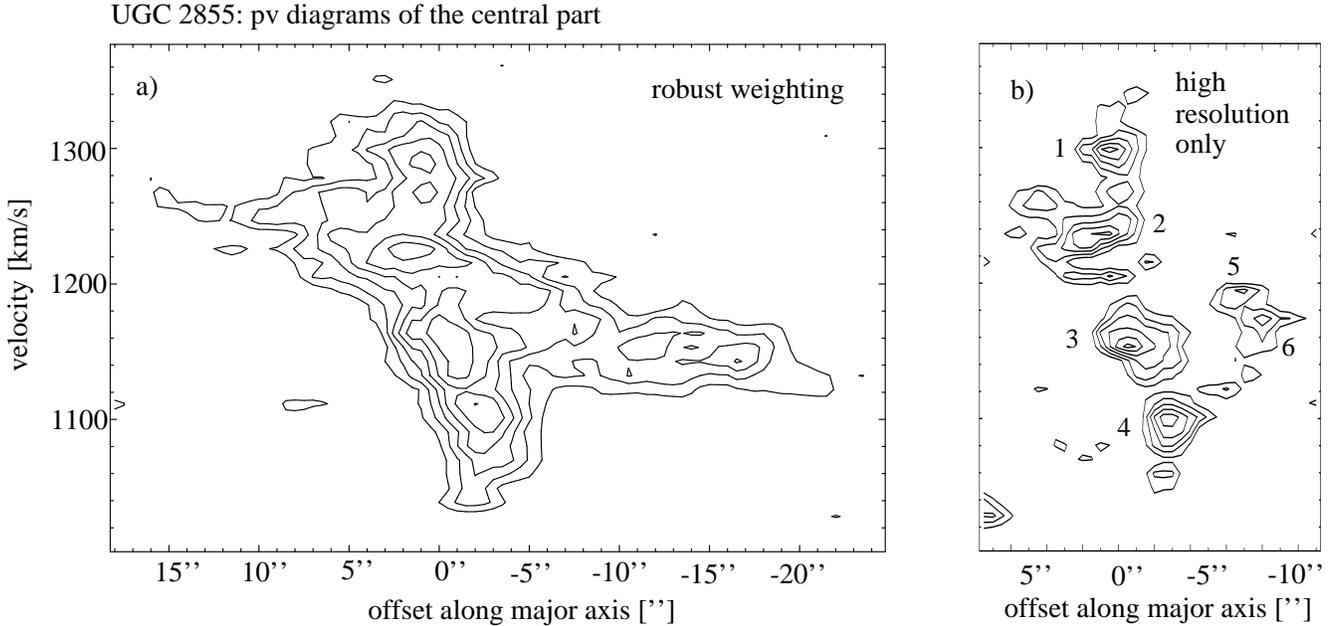}}
\caption{\label{2855pv_cen} Position-velocity diagrams of the central part
of the bar at higher resolution. Center and orientation of the cut are the 
same as in Fig.\,\label{2855pv} a. The robustly weighted data. The contours 
range from $2\sigma$ (0.86\,K \TB or 104\,mJy\,beam$^{-1}$) to 
4.38\,K in steps of $1.5\sigma$.
b) The high resolution track only. The lowest contour is again at $2\sigma$
(2.68\,K\,\TB or 124\,mJy\,beam$^{-1}$), and the contour step is $1\sigma$.
While the `main bar' is almost resolved out, the high velocity structure 
breaks up into a number of clumps, labels 1 --6.}
\end{figure*}

A position-velocity diagram of UGC\,2855 along the major axis of the
bar (Fig.\,6) shows solid body rotation in the bar out to its end 
at a radius of $\sim 4$\,kpc. Close to the center, extending to only a 
radius of 3$''$ or 290\,pc, there is another feature with a total
velocity width of almost 300\,\kms . Figure \ref{2855pv_cen} illustrates
the clumpiness of this structure at higher resolution.

An important question to be asked of a galactic bar is whether an Inner
Lindblad Resonance (ILR) exists and where it is located. The presence of an 
ILR affects the ease with which gas can reach the galactic nucleus: At an
ILR, material is often trapped in a circumnuclear structure and prevented 
from further infall. 

The general shape of the position-velocity diagram of UGC\,2855 could be 
modelled in terms of a bar with an ILR. Models like those
applied by Garc\'{\i}a-Burillo \& Gu\'{e}lin (1995) to NGC\,891 match the
pattern we see almost perfectly (see especially their Fig.\,8). The best
fit occurs if the bar of UGC\,2855 is seen almost side-on, i.e.\ along its 
minor axis, the angle also implied by the orientation of the bar parallel to 
the major axis of the galaxy. In this model, the `high velocity feature' 
represents gas on $x_2$ (anti-bar) orbits, while gas on bar-enforcing $x_1$ 
orbits causes the `main' feature.

{\it Despite the close match, caution is necessary in applying this
model directly to UGC\,2855.\/} The modelling of NGC\,891 assumed a weak
bar, while the bar of UGC\,2855 is strong. Strong bars, however, are 
not expected to have an ILR (Friedli 1998), and therefore $x_2$ orbits 
may not occur. A possible way out is an ILR that is located very close
to the nucleus, so that the $x_2$-to-$x_1$ coverage ratio remains very
small. In this case, long, gas-rich bars with an ILR may still be possible 
(Ishizuki 1997). From the structure we see, we can thus place the ILR
at an outermost radius of $\sim 3''$ (290\,pc). Still, even an ILR at $3''$
radius might betray itself by the presence of gas on $x_2$-orbits elongated
along the minor bar axis. Often, galaxies with an ILR show circumnuclear 
rings with emission maxima (`twin peaks') where the $x_1$ 
and $x_2$ orbits intersect and gas accumulates (see Kenney et al.\ 1992,
Downes et al.\ 1996, Sempere et al.\ 1995a for examples). No such structure
along the minor bar axis is visible in UGC\,2855. 
 
Thus, the bar of UGC\,2855 may {\em not have an ILR at all}. Since the
high velocity pattern shows no forbidden velocities (which are,
however, not expected for a viewing angle along the minor axis), it may 
also be a circularly rotating structure, i.e.\ a clumpy disk. This structure
may be fed efficiently by gas falling in along the bar, especially if
no ILR is present. 

The end of the bar at $\sim 4$\,kpc gives a firm lower limit to the 
location of the corotation radius (e.g.\ Elmegreen 1996).
If corotation occurs close to the bar end, as is often assumed and 
indicated by numerical simulations (e.g.\ Athanassoula 1992), a pattern 
speed of the bar of $\sim$\,5.2\,\kms\,arcsec$^{-1}$ or 53\,\kms\,kpc$^{-1}$
is derived. However, Combes \& Elmegreen (1993) argue that, especially 
in late-type galaxies, the length of the bar is determined by the scale 
length of the disk and corotation may be far outside the radius of the 
bar ends. 

The rotation curve along the major axis, which coincides with the bar axis, 
does not represent the circular rotation curve of the galaxy and cannot 
be used to determine $\Omega(r)$. Thus, a stringent determination of the 
location of the resonances has to await an evaluation of the potential 
of the galaxy, e.g.\ using a K-band image to obtain the mass distribution. 

\subsubsection{Velocity gradients}

The isovelocity contours along the bar of UGC\,2855 are almost
perpendicular to the bar major axis. However, some deviations are found.
Velocity differences, indicating streaming motions across the bar
measured parallel to the minor bar axis, are typically 20\,\kms\ in the
plane of the galaxy (assuming $i = 60^{\circ}$). There are 
variations, especially between the two sides of the galaxy (see 
Table\,\ref{vgrad}): The velocity gradient is generally higher in the
southeastern part of the bar, reaching a maximum of 46\,\kms\,kpc$^{-1}$
at an offset of 14$''$ from the center. 

\begin{table}
\caption{\label{vgrad} Streaming motions across the bar of UGC\,2855. 
$\Delta v$ gives the difference in velocity in the plane of the galaxy,
measured perpendicular to the bar axis.}
\begin{tabular}{lrr|lrr}
pos & $\Delta v$ & gradient & pos & $\Delta v$ & gradient \\
\arcsec  & \kms\ & $\frac{\rm km\,s^{-1}}{\rm kpc}$ & $''$ & \kms\ & 
$\frac{\rm km\,s^{-1}}{\rm kpc}$ \\ 
\hline
--30 & 18 & 15 & +30 & 17 & 12 \\
--22 & 19 & 20 & +22 & 29 & 22 \\
--14 & 14 & 12 & +14 & 36 & 46 \\
--6  & 12 & 9  & +6  & 29 & 27 \\
\hline
\end{tabular}
\end{table}

These velocity gradients perpendicular to the bar axis of UGC\,2855
are comparatively small and smooth. One could suspect that, if there is 
a very sharp discontinuity, even the high resolution of our interferometric
map might be insufficient to show it. However, the low dispersion we find 
in the region of the bar of UGC\,2855 where the streaming motions reach 
a maximum implies that there is indeed no smoothed out, unresolved shock 
feature hidden. 

A largely perpendicular orientation of the velocity contours with respect 
to the bar major axis is expected from simulations of gas streamlines in 
bar potentials if the bar is seen side-on. For this case, van Albada \& 
Roberts (1981) predict, however, a very sharp, spikelike discontinuity in 
velocity across the bar, indicating the presence of a shock. In a 
large number of model runs Athanassoula (1992) almost always finds sharp 
velocity jumps, connected to dust lanes, shocks and gas density enhancements, 
usually on the leading side of the bar, no matter whether the bar has an 
ILR or not. Interestingly, the {\em only model with no shock\/} in her sequence 
is the one having the {\em lowest central mass concentration (and no ILR)\/}. 
Since infall along a bar should increase the 
central mass with time, it seems likely that a situation like this prevails
early in the evolution of a bar. This agrees with the expectations of a 
linear theory of swing amplification for the initialisation of a bar, which
requires the bar to set out with no ILR (Toomre 1981).

The general shape of the velocity contours we observe agrees with the 
predictions of the ``no-shock''-model, especially on the more 
undisturbed northwestern side of the bar, where there are even hints 
of the expected `bulges' in the contours at offsets of about $-20''$ along 
the bar. 

The only velocity discontinuity found along the bar occurs very
locally, in a curved region at a major axis offset of about $+25''$, close 
to the bar end. Here, a jump of close to 80\,\kms\ over little more than 
$1''$ is seen, i.e.\ the discontinuity is unresolved and corresponds to a 
gradient of at least 600\,\kms\,kpc$^{-1}$. It is noteworthy that this 
region is located on the more disturbed side of the bar. 

\subsubsection{Properties of the inner structure}

If the inner high velocity structure is a circularly rotating disk, 
presumed to be aligned with the plane of the galaxy, i.e.\ seen at an 
inclination of 60$^{\circ}$, its dynamical mass will be $\sim 1.7 \cdot
10^9$\,M$_{\odot}$ (assuming a radius of 290\,pc and a circular velocity 
of 160\,\kms\ in the plane of the galaxy, estimated from 
Fig.\,\ref{2855pv_cen}). The total mass in the central region of UGC\,2855 
may thus be smaller than the mass in the inner 300\,pc radius 
of the Milky Way ($\sim 4 \cdot 10^{9}$\,M$_{\odot}$, e.g.\ McGinn et al.\ 1989). 

The H$_2$ mass from the high resolution data in this area
is $3.8 \cdot 10^8$\,M$_{\odot}$. This rises to $6 \cdot 10^8$\,M$_{\odot}$ 
if the naturally weighted map of the central region is evaluated. After the 
helium correction, $8.4 \cdot 10^8$\,M$_{\odot}$ or more 
than half of the dynamical mass in the central 3$''$ radius along the 
bar may thus be gaseous (for a lower gas mass, this percentage drops by
a factor of up to 8).

The clumpiness of the high velocity feature  (Fig.\,\ref{2855pv_cen} b) 
allows the identification of 6 molecular cloud complexes above the 
$3 \sigma$ level, labeled 1 to 6 in Fig.\,\ref{2855pv_cen}\,b. The complexes
5 and 6 show a clear drop in velocity and are likely to belong to the
main bar structure. 

The diameters of these complexes, which can only be resolved as distinct 
entities in a pv-diagram, are of order 300\,pc. With a velocity width of 
$\sim 30$\,\kms , they have virial masses of roughly $4 \cdot 10^7$M$_{\odot}$.
This matches the definition of Giant Molecular Associations (GMAs) coined by
Vogel et al.\ (1988). We find an H$_2$ (SCF) mass of typically 
$\sim 5 \cdot 10^7$\,M$_{\odot}$ for these structures, which may suggest that
they are bound objects.  

It is interesting to note that no intensity peak at the
center corresponds to the systemic velocity of 1200\,\kms . This explains the 
slight blueshift observed in the velocity field (Fig\,1\,b) at the
center. It might also suggest an explanation for the moderate star formation
even in the high surface density central region. The inner structure may
be better described as a torus with a central hole of radius $0.5''$ than as 
a centrally peaked disk. It it even conceivable that this structure is an
inner bar, possibly with its own pattern speed (and an ILR at 0.5$''$ radius).
There is, in fact, a hint of a slight misalignment of the major axis of 
the inner structure with respect to the main bar: both the high resolution 
$^{12}$CO and the $^{13}$CO data are matched better by PA of $125^{\circ}$ 
than by the $110^{\circ}$ determined for the large scale bar. 

\subsection{UGC\,2866}

\begin{figure}
\resizebox{\hsize}{!}{\includegraphics{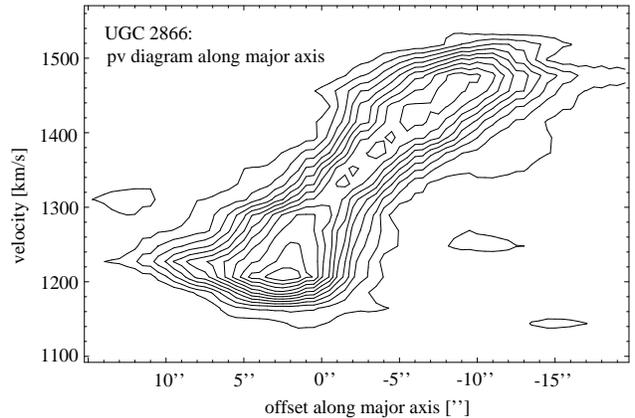}}
\caption{\label{2866pv} Position-velocity diagram along the major axis
for UGC\,2866. As is Fig.\,\label{25855pv}, the center of the cut coincides 
with the peak of the integrated CO emission, at $\alpha = 03^{\rm h} 45^{\rm m}
7.2^{\rm s};\ \delta = 69^{\circ} 56' 36''$. The lowest contour ($2\sigma$)
is at 0.28\,K\,\TB\ or 54\,mJy\,beam$^{-1}$. The distribution is dominated 
by one feature only.}
\end{figure}

The position-velocity diagram along the major axis of UGC\,2866 is displayed
in Fig.\,\ref{2866pv}. It confirms that all the molecular gas in this galaxy is
part of one kinematic structure. It is, however, not possible to decide whether
this structure is a bar or a circularly rotating disk. In the case of a 
rotating disk, inclined by 30$^{\circ}$, the dynamical mass is
$\sim 3.5 \cdot 10^{10}$\,M$_{\odot}$. However, if a circular disk was assumed,
the measured axis ratio of the CO structure of $\sim 0.4$ would imply an
inclination of $66^{\circ}$. Thus, the value of $\sim 30^{\circ}$ deduced
from the catalogued axis ratios seems too small. 
Alternatively, the observed discrepancy can be interpreted as pointing 
toward a bar morphology for the molecular structure. 
 
\subsection{Not all gaseous bars are equal: Evolutionary effects?}

From the rarity of continuous gas-rich bars it can already be inferred
that these phenomena have to be transient. This is true even though 
some caution is necessary: Only few galaxies have been fully mapped in
molecular gas, and thus bars rich in molecular gas may be more frequent 
than is apparent today (Turner 1996). We found one gas-rich system
with a long bar (UGC\,2855) and another system the morphology of which
is compatible with all gas being in bar-like structure (UGC\,2866)
among only three galaxies we have inspected so far. The ready success 
of our search for such systems suggests that such objects might not be 
extremely uncommon.

Evolutionary scenarios suggest that infall of matter along bars drives
a change from a later to an earlier galaxy type, since the flow concentrates
mass in the center, leading to a more pronounced bulge or nuclear region. 
UGC\,2855, classified as SBc, does not seem to have a large central 
mass concentration (though we do not yet have K-band data to derive the
mass distribution of the galaxy), so it may be at the beginning of the
concentration process. In this case, the bar may be very young, in line
with the notion that interaction with UGC\,2866 has recently triggered its
formation.  

\subsubsection{Gas and star formation properties: a comparison between 
UGC\,2855 and NGC\,7479}

Two of the longest gas-rich bars known, the one of NGC\,7479 and the one of
UGC\,2855 reported here, are very dissimilar objects. This is especially
obvious when one inspects the H$\alpha$ images: The bar of NGC\,7479 is
clearly outlined in H$\alpha$ emission, indicating vigorous star formation
all along it (Martin \& Friedli 1997, Quillen et al.\ 1995). Dust lanes at
the leading edge of the bar are widely regarded as indicators of a
shock front (e.g.\ Athanassoula 1992). In contrast, 
the bar in UGC\,2855 is only barely visible in H$\alpha$ (Fig.\,\ref{halpha} 
a). It is interesting to note that the part of the bar southeast of the 
nucleus, which is weaker in (interferometrically detected) CO, {\em 
but more disturbed in velocity\/}, is the region of the bar where 
faint H$\alpha$ can be most clearly detected. 

May the gas in the bar of UGC\,2855 be in a more quiescent state than in
the bar of NGC\,7479? Clearly, the degree of star formation as indicated by
H$\alpha$ emission varies over a wide range in barred galaxies (Martin
\& Friedli 1997).

\begin{table}
\caption{\label{ratios} $^{12}$CO/$^{13}$CO total integrated line ratios for
UGC\,2855 and, for comparison purposes, NGC\,7479 and typical galactic ratios. 
$1\sigma$ errors are given in parantheses.}
\begin{tabular}{l|l|r}
Position & Telescope & ratio \\
\hline 
UGC\,2855 center & OVRO & 9.5(2.0) \\
UGC\,2855 center & OSO 20m & 10.0(1.0) \\
UGC\,2855 $+45''$ & OSO 20m & 8.2(5.0) \\
UGC\,2855 $+30''$ & OSO 20m & 7.9(1.6) \\
UGC\,2855 $-30''$ & OSO 20m & 5.1(4.0) \\
UGC\,2855 $-45''$ & OSO 20m & 5.0(1.3) \\
NGC\,7479 center & OVRO & 15 -- 35 \\
NGC\,7479 along bar & OVRO & $> 20$ \\
NGC\,7479 bar ends & OVRO & 6 -- 10? \\
galactic disk & --- & $\sim 6^{\rm a)}$ \\
centers of `normal' galaxies & --- & $13(6)^{\rm b)}$ \\
\hline 
\end{tabular} \\
a) Polk et al.\ 1988; b) Aalto et al.\ 1995 \\
\end{table} 

$^{12}$CO/$^{13}$CO line ratios may give a hint here, since they allow
insight into gas properties: In the cool ISM of the Galatic disk, where
the $^{12}$CO $1 \to 0$ line has a high optical depth, this ratio is
$\sim 6$ (Polk et al.\ 1988). Centers of normal galaxies (including,
at least on scales averaged over a large part of the Central Molecular
Zone, our own Galactic center region) typically have $^{12}$CO/$^{13}$CO 
ratios ranging from 10 -- 20.
A high ratio is ($\ga 15$) can indicate higher gas kinetic temperatures, 
the presence of diffuse molecular gas (e.g.\ Aalto et al.\ 1995), or, 
alternatively, higher gas densities, though, from non-LTE models, this 
effect alone cannot account for ratios exceeding $\sim 15$. 
These relations can only serve as a rough guide, since they do not 
include abundance effects, and a multi-transition study including more 
than two lines is needed for a proper excitation analysis that allows us 
to distinguish between the possibilities listed above. Still, a 
varying $^{12}$CO/$^{13}$CO line ratio is a clear indicator of changes 
in the gas properties. 

In NGC\,7479, large variations are found, with the ratio exceeding 20
along the bar, where the OVRO interferometer does not detect $^{13}$CO,
and a very variable (15 -- 35) ratio in the center (Aalto et al.\ 1998).
Our interferometric and complimentary single dish measurements, which have 
detected $^{13}$CO in several positions along the bar, indicate a more 
constant line ratio of just below 10 along the bar in UGC\,2855, 
with only a slight trend toward an increase (to 10) in the center. 
This might mean that the amount of diffuse or hot gas in the bar of UGC\,2855
is lower than in NGC\,7479 (see Table\,\ref{ratios}). The narrow line 
widths found in the bar of UGC\,2855 support this notion. Of course, 
caution is required when interpreting these line ratios, since the 33$''$ 
beam of the OSO single dish telescope picks up emission from the spiral 
arms. Still, the excellent agreement of the ratios between the 
interferometer and the single dish telescope in the center  implies 
that not much diffuse emission missed by the interferometer is present there. 
This is confirmed by the flux comparison between interferometer and single 
dish data (Sect.\ 3.1.2). 

It is suggestive (though only marginally significant) that the 
OSO $^{12}$CO/$^{13}$CO line ratios in the northwestern part of the bar,
which shows a more regular velocity field than the southeastern part, 
are slightly lower, possibly indicating even more quiescent, `disk-like' 
gas properties. Along the same line of argument, more flux seems to be 
missing from the interferometer map in the southwestern bar, and more
H$\alpha$ emision is detected, indicating a larger amount of diffuse gas 
and less quiescent conditions in this region. 

We would expect shocks along the bar to trap the diffuse gas and prevent
it from moving into the center more effectively than the dense gas. Thus,
the observation of lower $^{12}$CO/$^{13}$CO line ratios along the bar of
UGC\,2855 than in the bar of NGC\,7479 ties in very well with the 
indication of less or no shocks along its bar derived from its velocity 
field (Sect.\ 4.1.2). 

The velocity field of NGC\,7479 is distinctly different from the regular
field of UGC\,2855, where the contours are almost perpendicular to the bar 
major axis. In contrast, the velocity field in NGC\,7479 
appears far more disturbed. The velocity contours close to the center of 
NGC\,7479 at high resolution are almost parallel to the bar axis 
(Aalto et al.\ 1999 in prep.\, the field shown by  Sempere et al.\ (1995b) 
based on single dish observations does not show this clearly). Part of the 
difference between the two galaxies can be attributed to different viewing 
geometries, since the angle between the bar axis and the line of nodes in 
NGC\,7479 ($\sim 20^{\circ}$) suggests that the bar is not seen entirely 
side on. Indeed, early optical work on the velocity fields of barred galaxies 
(Pence \& Blackman 1984) as well as the numerical simulations (van Albada 
\& Roberts 1981) already clearly indicate that `S shaped' velocity contours, 
which are partially parallel to the bar axis, are best seen in galaxies 
where the orientation of the bar is close to 45$^{\circ}$ with respect to 
the line-of-sight. 

Still, all evidence, both dynamical and excitational, points toward there
being no large-scale shock present along the bar of UGC\,2855, 
while the discontinuities in the velocity structure of NGC\,7479 are 
consistent with a shock. This is also indicated by a shock-like 
feature discovered at the leading edge of the bar in NIR color maps by
Quillen et al.\ (1995). 

In the future, an optical image of UGC\,2855 that allows the detection
(or exclusion) of the presence of dust lanes will be helpful to finally decide
the question of whether the gas in the bar of this galaxy indeed manages
to escape being shocked.

The FIR $S_{100}/S_{60}$ color index is routinely used to indicate dust
(color) temperatures. The value of 2.6 found for UGC\,2855 suggests 
$T_{\rm dust} \sim 30 - 33$\,K, assuming a spectral emissivity index of
1.5 and 1.0 for the lower and upper temperature limits, respectively, and
applying color corrections. $T_{\rm dust}$ for NGC\,7479 may be slightly 
higher, with corresponding values $\sim 34 - 37$\,K. Estimating dust 
temperatures using the 60\,$\mu$m flux is problematic since the 60\,$\mu$m 
flux can have a large contribution from stochastically-heated dust grains,
thereby over-estimating the temperature of the thermal-equilibrium 
grains that dominate the mass of the dust (see e.g.\ D\'esert et al.\ 1990,
Wall et al.\ 1996). This effect is likely to be most pronounced in large 
spirals like UGC\,2855 and NGC\,7479 where the IRAS flux encompasses more 
than the nucleus and the bar. Still, these color temperatures hint at more star 
formation activity in NGC\,7479. Interestingly, UGC\,2866 has, again under 
the conventional assumptions, a high $T_{\rm dust}$ of $39 - 44$\,K. This 
appears to be consistent with the high degree of star formation activity 
indicated by the H$\alpha$ emission.    

\subsubsection{The evolutionary state of the bar of UGC\,2855}

Is it possible to place the bar of UGC\,2855 in an evolutionary scheme? 
Martinet \& Friedli (1997) use the IRAS color index $\log(S_{25}/S_{100})$
as an indicator of star formation activity, relate it to bar strength and 
length and calculate 3-dimensional self-consistent models of bar evolution
to predict the variation of these parameters with time. In their scheme, 
UGC\,2855, having $\log(S_{25}/S_{100}) = -1.26$, (barely) belongs to a
the group of galaxies with less pronounced star formation. Since the bar
is both long and strong, this places it in either an early, preburst stage  
($t_{\rm bar} \la 400$\,Myr) or in a late, postburst state ($t_{\rm bar} \ga
2000$\,Myr). The central-to-bar H$\alpha$ emission ratio suggests an early
evolutionary stage in the models of Martin \& Friedli (1997). However, these
models are still unable to explain all observed bar properties and differ
significantly depending on the initial parameters, e.g.\ the star formation
efficiency and the mechanical energy released into the ISM. 

On the side of the observations, it has also to be kept in mind that the large
IRAS beam may pick up significant emission from the gas-rich spiral arms
of UGC\,2855, and that the H$\alpha$ activity is generally low in the bar, 
making the ratio uncertain. Still, these indicators together with the
high gas content of the bar and the moderate mass concentration in the center
let us speculate that the bar in UGC\,2855 may be young. Then, this
presently inconspicious galaxy has the potential to become a spectacular 
starburst in the future.

\section{Conclusions}

The main results of our high resolution study of the stellar bar region of 
the SBc galaxy UGC\,2855 and of its companion UGC\,2866 are:

\begin{enumerate}
\item The interferometric image of the $^{12}$CO $1 \to 0$ transition of
UGC\,2855 reveals a continuous molecular bar of length $\sim 8$\,kpc. 
UGC\,2855 thus becomes one of only a few galaxies known to have dense 
gas distributed all along its stellar bar. The molecular mass detected 
in the bar is $\sim 3\ 10^9$\,M${_\odot}$, if a standard CO-H$_2$ conversion
factor is assumed. The intensity distribution appears strongly centrally
peaked.  

\item The velocity field of the gas along the bar is dominated by solid
body rotation. Only slight perturbations, indicative of streaming motions 
across the bar, are found. The smooth changes and lack of discontinuities 
in the velocity field together with the narrow line widths suggest that the 
gas along the bar of UGC\,2855 does not experience strong shocks. 

\item The $^{12}$CO/$^{13}$CO ratio along the bar and in the center is
almost constant at 5 -- 10. This ratio, which is typical for the disks (5)
and centers (10) of normal galaxies, also indicates quiescent conditions
along the bar. 

\item Only weak H$\alpha$ emission is detected along the bar of UGC\,2855.
The emission is stronger in the southeastern part of the bar, where the
velocity field of the molecular gas is more disturbed. The center is clearly
detected in H$\alpha$, but the strongest H$\alpha$ peak is associated with 
a spiral arm. Higher extinction in the nucleus might change this ratio,
but even then the central region would not dominate in H$\alpha$ as it
does in many starburst galaxies. Thus, while there is some central star 
formation activity, UGC\,2855 is not (yet) a starburst galaxy, as is 
also indicated by the IRAS $S_{25}/S_{100}$ color index. 

\item A high velocity structure dominates the inner $\sim 3''$ radius of 
the bar. This structure is responsible for the strong central intensity 
peak. It breaks up into a number of possibly virialised GMAs, the velocity 
of none of which coincides with the systemic velocity. This structure
may be a clumpy disk, a torus or even an inner bar, in all cases fed by 
gas infalling along the main bar. In the simple case of a rotating disk, 
more than 50\% of the dynamical mass ($1.7 \cdot 10^9$\,M${_\odot}$, 
which is less than the mass in the corresponding inner region of the Galaxy) 
may be gaseous. 

\item In models, shocks along a bar can be avoided if the central mass 
concentration is low. Then, the bar has no ILR. Even though the velocity
structure seen can also be explained in terms of a bar with an ILR, this
is an attractive model for UGC\,2855 since it naturally explains the
quiescent gas along found along the bar, as well as the absence of any 
feature along the bar minor axis expected from gas on $x_2$ orbits. 

\item In contrast, the gas-rich bar of NGC\,7479 shows a disturbed velocity
field, varying $^{12}$CO/$^{13}$CO line ratios and strong H$\alpha$ 
emission both along the bar and in the center. This demonstrates that
even among the small group of galaxies with gas-rich bars large 
differences are found.

\item In evolutionary terms, we suggest that the bar of UGC\,2855 is
a very young object, where the process of central concentration is 
just starting. Then, this galaxy is likely to develop into a starburst in 
the future, a state that NGC\,7479 has already reached.  

\item UGC\,2866, the companion of UGC\,2855, shows strong CO emission 
(corresponding to $\sim 2 \cdot 10^9$\,M${_\odot}$ with a standard conversion
factor) in a continuous, possibly bar-like feature of about 1.5\,kpc radius. 
Both the FIR colors and the very strong H$\alpha$ emission concentrated in 
the inner 300\,pc indicate that this galaxy is already experiencing a 
strong starburst. 

\item The activity in both UGC\,2866 and UGC\,2855 may have been triggered by
interaction between the two galaxies. The response time in the small, compact 
galaxy, UGC\,2866, may be shorter than in the large, open spiral UGC\,2855,
thus explaining the different level of activity seen at present.

\end{enumerate}

\acknowledgements{We thank John Black for first drawing our attention to
a pair of galaxies that only seemed inconspicious. We are very grateful to 
Martin Altmann for taking the H\,$\alpha$ image at Hoher List and helping
with its processing. We also gratefully acknowlegde the support of the 
staff of OVRO and OSO. The OVRO mm-array is supported in part by NSF grant
AST 9314079 and the K.T.\ and E.L.\ Norris Foundation.}

\end{document}